\documentclass[runningheads]{llncs}
\usepackage{epsfig}
\usepackage{graphicx}
\usepackage{amsmath}
\usepackage{amssymb}

\usepackage{subfigure}
\usepackage{multirow}
\usepackage{array}
\usepackage{color}
\usepackage{tabu}
\usepackage{longtable}

\usepackage[breaklinks=true,letterpaper=true,colorlinks,bookmarks=false]{hyperref}
\newcommand{\etal}{\textit{et al}. }
\newcommand{\ie}{\textit{i}.\textit{e}. }
\newcommand{\eg}{\textit{e}.\textit{g}.}

\begin{document}
\title{One Click Lesion RECIST Measurement and Segmentation on CT Scans}
%

\author{Youbao Tang\inst{1} \and
Ke Yan\inst{1} \and
Jing Xiao\inst{2} \and
Ronald M. Summers\inst{1}}
%
\authorrunning{Youbao Tang et al.}
%
\institute{Imaging Biomarkers and Computer-Aided Diagnosis Laboratory, Radiology and Imaging Sciences, National Institutes of Health Clinical Center, Bethesda, MD 20892-1182, USA \\ \email{youbao.tang@nih.gov, rms@nih.gov} \and
Ping An Insurance Company of China, Shenzhen, 510852, China}

\maketitle              

\begin{abstract}
In clinical trials, one of the radiologists' routine work is to measure tumor sizes on medical images using the RECIST criteria (Response Evaluation Criteria In Solid Tumors). However, manual measurement is tedious and subject to inter-observer variability. We propose a unified framework named SEENet for semi-automatic lesion \textit{SE}gmentation and RECIST \textit{E}stimation on a variety of lesions over the entire human body. The user is only required to provide simple guidance by clicking once near the lesion. SEENet consists of two main parts. The first one extracts the lesion of interest with the one-click guidance, roughly segments the lesion, and estimates its RECIST measurement. Based on the results of the first network, the second one refines the lesion segmentation and RECIST estimation. 
SEENet achieves state-of-the-art performance in lesion segmentation and RECIST estimation on the large-scale public DeepLesion dataset. It offers a practical tool for radiologists to generate reliable lesion measurements (\ie segmentation mask and RECIST) with minimal human effort and greatly reduced time.

\end{abstract}
\section{Introduction}
Lesion segmentation and measurement from computed tomography (CT) scans are important tasks in oncology image analysis. They are useful in quantitative assessment of the disease progression and therapy response. Generally, radiologists scan through the CT image to find lesions, and then measure their sizes using the RECIST (Response Evaluation Criteria In Solid Tumors) criteria  \cite{eisenhauer2009new}. They usually do not segment lesions, even though segmentations might prove helpful, due to time constraints. 
With these measurements, follow-up and quantitative analysis of tumor extents could be performed to provide valuable information for treatment planning and tracking. Such manual annotation processes are highly tedious and time-consuming, so it motivates many researchers to develop techniques to automate these processes. 

With the advent of deep learning, applications of medical image analysis \cite{WANG2017172,cai2018accurate,li2018h,tang2020automated,pelvic_yirui,tang2019uldor,agarwal2020weakly,tang2019tuna,yan2019mulan,jin2018ct,tang2018semi,yan20183d,li2019mvp,zlocha2019improving,tao2019improving,tang2019ct,tang2019xlsor,zhu2020cross,tang2020e2net,chen2020anatomy,tang2019abnormal} using deep learning have dramatically increased in the last few years. Many previous work focus on a single lesion type over a specific body part, \eg, lung nodule segmentation \cite{WANG2017172}, lymph node segmentation \cite{tang2019ct} and liver tumor segmentation \cite{li2018h}. However, radiologists in their daily work need to find and measure all kinds of lesions, so single-type detection and segmentation algorithms are not scalable in practice. Recently, a large-scale DeepLesion dataset \cite{yan2018deeplesion} covering different types of lesions over the entire body was released, which inspired techniques on universal lesion analysis. While universal lesion detection has received much attention \cite{yan20183d,tang2019uldor,li2019mvp,zlocha2019improving,tao2019improving,yan2019mulan}, segmentation and RECIST estimation were not sufficiently studied.
To accurately locate the lesion to segment or measure, existing work \cite{cai2018accurate,tang2018ct,tang2018semi,agarwal2020weakly} require strong human guidance information, \eg, drawing a bounding box to cover the lesion region that should not be too tight or too large. This constraint is not user-friendly and increases time cost.

To overcome these problems, we propose a unified framework, SEENet, for joint lesion segmentation and RECIST estimation over the whole human body with only one click guidance. 
Unlike previous work \cite{cai2018accurate,tang2018ct,tang2018semi,agarwal2020weakly} that ask radiologists to draw a bounding box to indicate the lesion of interest (LOI), this work only requires a single click, which is more convenient and efficient, without the need to carefully control the box size when drawing. More importantly, SEENet achieved better accuracy and robustness than \cite{cai2018accurate,tang2018ct,tang2018semi,agarwal2020weakly}.
SEENet consists of two main components. The first component (named MR-CNN) is responsible for simultaneous LOI extraction, initial lesion segmentation and RECIST estimation. We improved the Mask R-CNN \cite{he2017mask} by adding a new branch for predicting the endpoints' heatmaps of RECIST. One-click guidance provided by users is required as the input of MR-CNN. 
For the second component, a model (named ARU-Net) is built by combining the advantages of U-Net \cite{ronneberger2015u}, ResNet \cite{He_2016_CVPR} and atrous spatial pyramid pooling module (ASPP) \cite{chen2018deeplab} for lesion segmentation and RECIST estimation refinement. ARU-Net is able to learn highly discriminative features that consider multiscale contextual information from image patches obtained according to the outputs of MR-CNN for full-size pixel-wise prediction. 

The proposed SEENet is trained and evaluated on the large-scale DeepLesion dataset \cite{yan2018deeplesion}. The coordinates of RECIST measurements provided in DeepLesion are use as supervisory information to learn the lesion segmentation task in a weakly-supervised way, so manual mask annotation for training is not needed. Extensive experimental results demonstrate that the proposed SEENet achieves the state-of-the-art performance in both universal lesion segmentation and RECIST estimation tasks, and MR-CNN without using the click information also achieves the competitive performance of fully automatic lesion detection. Overall, this work tries to solve a clinically important problem. It allows clinicians to easily control which lesion to segment and measure by one click and the rest is done accurately and automatically.

The main contributions of this work can be summarized as follows: 1) This work provides a solution to an important clinical task of lesion measurement annotation and segmentation. 2) A network based on Mask RCNN is built to extract lesion of interest, estimate RECIST measurement and segment lesion simultaneously. 3) A network that can learn multi-scale contextual information of lesions and perform full-size prediction is built to refine lesion segmentation and RECIST estimation results.

\section{Methodology}
\label{mrcnn}
\begin{figure*}[t!]
  \centering
  \includegraphics[width=\linewidth]{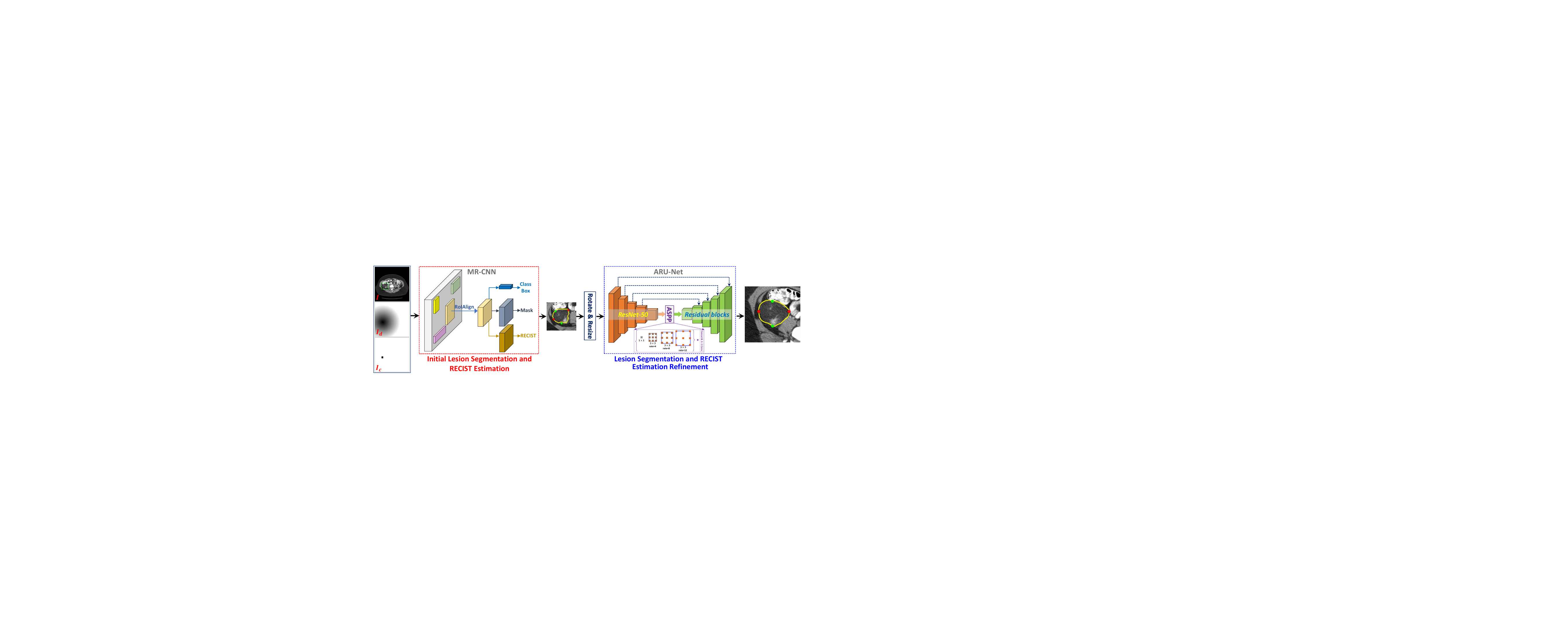}
  \caption{Overview of the proposed SEENet. Yellow curves indicate the boundaries of the segmented lesions. Red and green spots indicate the endpoints of the long and short axes of the estimated RECIST measurement, respectively.}
  \label{fig:framework}
\end{figure*}

Given a CT scan with lesions, we propose a unified framework to segment the lesion region and estimate the RECIST measurement accurately and simultaneously with one-click guidance from human. Fig. \ref{fig:framework} shows an overview of the proposed SEENet. It consists of two main components. The first one is designed based on an improved Mask R-CNN \cite{he2017mask} for initial lesion segmentation and RECIST estimation. The second one is a U-Net \cite{ronneberger2015u} like architecture for lesion segmentation and RECIST estimation refinement. The motivation and detail of their design will be described in this section.

\textbf{Initial Lesion Segmentation and RECIST Estimation.}
Mask R-CNN \cite{he2017mask} is able to detect and segment objects at the same time, and it has been successfully applied for lesion detection in \cite{tang2019uldor}. With these inspirations, we propose a model (named MR-CNN) based on Mask R-CNN for simultaneous lesion segmentation and RECIST estimation. To achieve this goal, we need to first locate the lesion of interest (LOI), then predict lesion segmentation and RECIST estimation for the LOI.

Although previous works \cite{tang2019uldor,yan20183d,li2019mvp,zlocha2019improving,tao2019improving,yan2019mulan} contribute many efforts to improve the detection performance, there still exist many false positives in their lesion detection results. Our goal is to let radiologists segment and measure the lesion of their interest and not to distract them with false positive lesion detection results, so we need some manual input to indicate the LOI. Existing approaches \cite{cai2018accurate,tang2018ct,tang2018semi,agarwal2020weakly} require radiologists to manually draw bounding boxes to obtain LOIs for their tasks. 
Compared with \cite{cai2018accurate,tang2018ct,tang2018semi,agarwal2020weakly}, a faster and easier way adopted in this work is to click around or inside the lesion region once. To effectively inject the one-click guide information into the model, a click image $I_c$ and a distance transform image $I_d$ are generated and truncated at 255 from the click point $\mathbf{p}$, where $I_d$ is defined as $I_d(\mathbf{q})=\lVert \mathbf{q}-\mathbf{p}\rVert_2$. Therefore, a three-channel image can be constructed as the input by concatenating the original 2D CT slice $I$, $I_c$ and $I_d$, as shown in Fig. \ref{fig:framework}.

The original Mask R-CNN has two branches, one for detection and the other for segmentation. We hope the network to also predict the RECIST measurement, which consists of two lines: one measuring the longest axis of the lesion and the second measuring its longest perpendicular axis in the axial plane, see Fig. \ref{fig:result} for examples. Extending Mask R-CNN, the proposed MR-CNN outputs a RECIST estimation result in parallel by adding a new branch for predicting the heatmaps of four keypoints (\ie the four endpoints of the RECIST axes), as shown in Fig. \ref{fig:framework}. Therefore, MR-CNN contains a backbone and three head branches for LOI extraction (classification and regression), mask prediction, and RECIST estimation over each LOI. Following the framework of \cite{yan2019mulan}, we remove the last dense block and transition layer of DenseNet-121 \cite{huang2017densely} and use it as the backbone in this work. The box recognition and mask prediction branches are the same as Mask R-CNN. The added RECIST estimation branch has a similar structure as the mask prediction branch except with a four-channel output.

Like previous work \cite{tang2019uldor,yan20183d,li2019mvp,zlocha2019improving,tao2019improving,yan2019mulan}, MR-CNN can also be employed for fully automatic lesion detection when without using the click information as input. Since the three branches of MR-CNN are jointly trained, we find that the added RECIST estimation branch can enhance lesion detection performance. This will be validated in the section of experiments.

\textbf{Lesion Segmentation and RECIST Estimation Refinement.}
The initial mask and RECIST prediction is not sufficiently accurate and fine-grained, since the output size of mask and RECIST branches is $28\times 28$, which is too small to get accurate pixel-wise prediction. Therefore, we design a refinement model according to the following three principles: 1) performing full-size prediction, 2) using a strong backbone for feature extraction, and 3) considering rich contextual information at multiple scales due to the size variation of lesions. After investigating previous pixel-wise prediction approaches, we find that 1) the U-Net\cite{ronneberger2015u} is well designed for full-size prediction and has been successfully used in many medical image segmentation tasks, 2) ResNet-Pose \cite{xiao2018simple} achieves good human pose estimation performance due to its powerful backbone (\ie ResNet \cite{He_2016_CVPR}), and 3) the atrous spatial pyramid pooling module (ASPP) \cite{chen2018deeplab} is able to learn multiscale feature representations with rich contextual information. Inspired by these works, we propose ARU-Net with a U-Net like architecture, using ResNet in the encoder and decoder, and ASPP as the junction between them for lesion segmentation and RECIST estimation refinement, as shown in Fig. \ref{fig:framework}.

Specifically, the first five blocks of ResNet-50 are used as the encoder of U-Net and the stride of the first convolutional layer is set as 1 instead of 2 to keep more spatial information in the output feature maps. Then, an ASPP module with 256 output channels and four different dilation rates (\ie 1, 4, 8 and 12) follows the encoder, and their outputs are concatenated to form the multiscale contextual feature maps as an input of the decoder, as shown in Fig. \ref{fig:framework}. The decoder of U-Net consists of four upsampling blocks and one output block. Each upsampling block has a linear upsampling layer and two residual blocks. The output channel sizes of upsampling blocks are 512, 256, 128 and 64, respectively. The output block has two branches in parallel, one for mask prediction and the other for RECIST estimation.

Based on the results produced by the trained MR-CNN model (including a detected box, a predicted mask and an estimated RECIST), an adjusted LOI can be obtained as the input of ARU-Net with the following steps: 1) rotating the CT scan according to the center and the long axis' slope of the estimated RECIST; 2) cropping a square sub-image whose center is the predicted mask and whose width is two times the extent of the detected box's long side, so that sufficient visual context is preserved for ARU-Net training; 3) resizing the cropped image to $256\times 256$.

\textbf{Model Optimization.}
As done by the Mask R-CNN \cite{he2017mask}, a multi-task loss on each sampled RoI is defined as $L^1=L^1_{cls}+L^1_{box}+L^1_{mask}+L^1_{recist}$ for MR-CNN training. The first three terms have the same definitions as the ones in \cite{he2017mask}. $L^1_{recist}$ is the mean squared error (MSE) loss. Another multi-task loss is defined as $L^2=L^2_{mask}+L^2_{recist}$ for ARU-Net training, where both $L^2_{mask}$ and $L^2_{recist}$ are the MSE loss. For $L^1_{mask}$ update, the pseudo masks of lesions are constructed based on GrabCut \cite{grabcut} from RECIST annotations following \cite{tang2018semi}. For $L^2_{mask}$ update, we use the prediction of MR-CNN to refine the potentially noisy ground-truth. Specifically, the intersections of the segmentation results produced by MR-CNN and the pseudo masks are labeled as lesion and their differences are labeled as uncertain regions that will be ignored during training. For $L^1_{recist}$($L^2_{recist}$) update, ground-truth keypoint heatmaps consist of four 2D \mbox{Gaussian} maps (with a standard deviation of 1(3) pixel(s)) centered on the endpoints of RECIST annotations. Both MR-CNN and ARU-Net are implemented in PyTorch. To mimic the radiologists' click behavior during training, we randomly sample the points from the dilated pseudo masks with five pixels for generating clicks. The \mbox{ImageNet} pre-trained models are used for weight initialization. MR-CNN and ARU-Net are separately trained. We first train MR-CNN using SGD with an initial learning rate of 0.004 for 8 epochs and decay it by 0.1 after 4 and 6 epochs, and then train ARU-Net using SGD with an initial learning rate of 0.01 for 150 epochs and decay it by 0.1 after every 50 epochs. Regular image transformations including scaling, rotation, and translation were used for data augmentation.

\section{Experiments}

\textbf{Dataset.}
The DeepLesion dataset \cite{yan2018deeplesion} is composed of $32,735$ CT lesion images annotated with RECIST measurements from $10,594$ studies of $4,459$ patients. It contains a variety of lesions throughout the body such as lung nodules, liver lesions, and enlarged lymph nodes. Following the previous work \cite{cai2018accurate}, $1,000$ lesion images are randomly selected from 500 patients and manually segmented as a test set for quantitative evaluation. The remaining patients' images serve as a training set (90\%) and a validation set (10\%).

\textbf{Evaluation Criteria.}
For lesion segmentation, the pixel-wise precision, recall, and dice similarity coefficient (Dice) are used for performance evaluation.
For RECIST estimation, to quantify how our approach measures against the radiologists' annotations, we compare the RECIST annotations and those of our method by computing the mean and standard deviation of differences between their diameter lengths.
As introduced in Section \ref{mrcnn}, MR-CNN can be used for fully automatic lesion detection. Therefore, we also evaluate the lesion detection performance using the free-response receiver operating characteristic (FROC), following previous works \cite{tang2019uldor,yan20183d,li2019mvp,zlocha2019improving,tao2019improving,yan2019mulan}. A detection result is considered as correct when the IoU between the predicted bounding box and the real bounding box is larger than 0.5.

\begin{figure*}[t!]
	\begin{minipage}[b]{1.0\linewidth}
		\centering
		\includegraphics[width=0.99\linewidth]{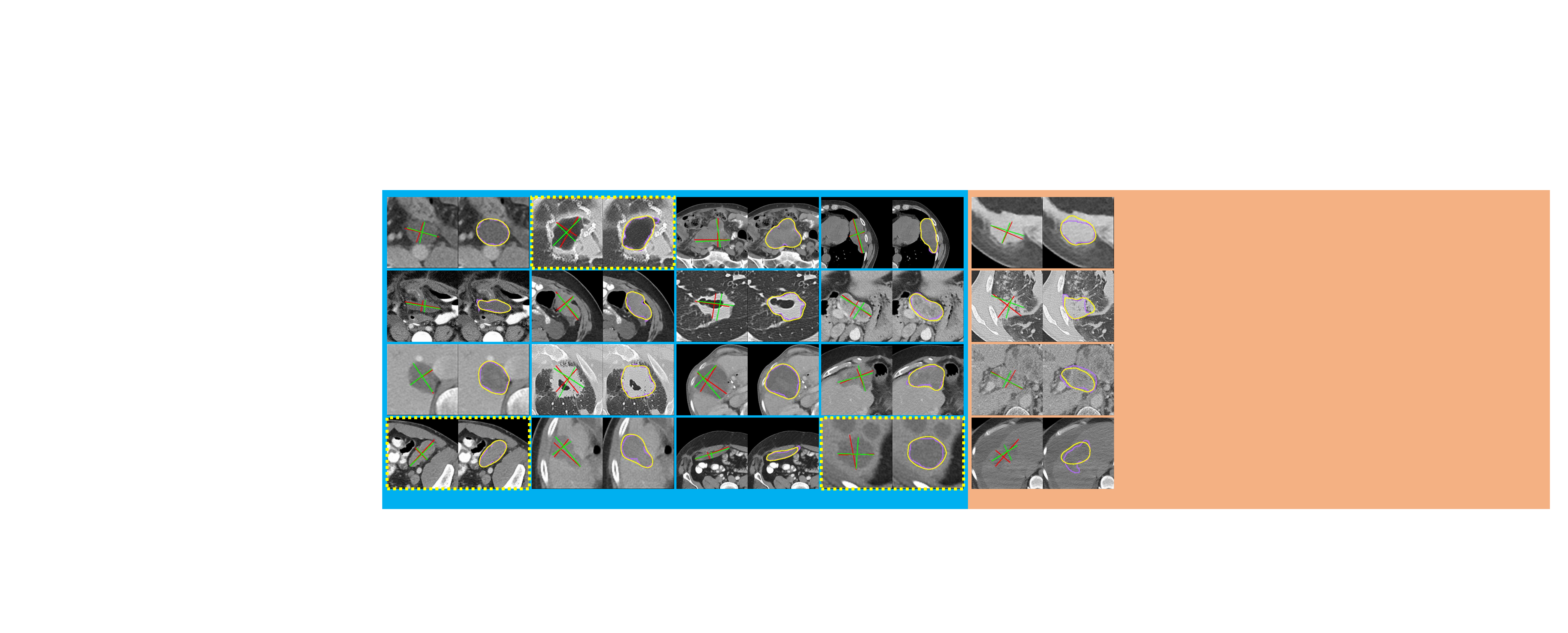} \\
	\end{minipage}
    \begin{minipage}[b]{0.8\linewidth}
		\centering
		\centerline{(a)}\medskip
	\end{minipage}
	\begin{minipage}[b]{0.18\linewidth}
		\centering
		\centerline{(b)}\medskip
	\end{minipage}
	\caption{Visual examples of results produced by SEENet. For clear visualization, the lesion segmentation and RECIST estimation results are shown separately. Therefore, each example includes two images. The left one presents the estimated RECIST (the green cross) and the RECIST annotation labeled by radiologists (the red cross). The right one presents the lesion segmentation result (the yellow curve) and the real lesion mask (the pink curve). The yellow dashed boxes indicate some cases where the manual RECISTs are imperfect, but SEENet can fix them (see text). (a) Sixteen good quality examples, (b) two over-estimated (top) and two under-estimated examples (bottom). All images in this figure are cropped from original CT scans according to the results produced by MR-CNN. Best viewed in color.}
	\label{fig:result}
\end{figure*}

\textbf{Qualitative Results.}
Fig. \ref{fig:result} shows several visual examples of the results produced by SEENet. From Fig. \ref{fig:result}, all lesion regions are well cropped, suggesting that SEENet can successfully extract accurate lesion of interests (LOIs) with less human effort (\ie one manual click) instead of manually drawing bounding boxes to include the lesion regions as done by previous work \cite{cai2018accurate,tang2018ct,tang2018semi,agarwal2020weakly}. Some good quality examples are given in Fig. \ref{fig:result}(a), where the automatic lesion segmentation results (the yellow curves)  are very close to the manual segmentations (the pink curves) and the estimated RECISTs can well provide the information of lesions' long and short diameters. As demonstrated by \cite{tang2018semi}, there is a large variation between RECIST annotations from different radiologists. Therefore, it's easy to understand that the endpoints of the estimated RECISTs are different from the ones of manual annotations. We also find that our estimation results are close to each other even the clicks are given at different positions. Sometimes, although the manual annotations cannot well touch the boundaries of the lesion regions, our estimated results can fix them (seeing the examples indicated with yellow dashed boxes). Two over(under)-estimated results are given in the top (bottom) of Fig. \ref{fig:result}(b). In these cases, the boundaries of lesions are highly blurred, which makes SEENet fail to extract highly discriminative features from them to distinguish the lesions from their surrounding regions, especially for lesion segmentation, since only weak pixel-wise labels are used for training. These qualitative results intuitively demonstrate that the proposed framework SEENet can well perform the lesion segmentation and RECIST estimation simultaneously with one-click guide information during inference.

\begin{table}[t!]
	\begin{center}
		\caption{Results of lesion segmentation and RECIST estimation. For lesion segmentation, the mean and standard deviation of pixel-wise recall, precision and Dice score are reported. For RECIST estimation, the mean and standard deviation of the differences of diameter lengths (mm) between radiologist RECIST annotations and those obtained by different methods are reported.}
		\label{tab:recist}
		{
			\scriptsize
			
			\begin{tabular}{|@{}*{1}{m{3.5cm}<{\centering}@{}}|@{}*{1}{m{1.7cm}<{\centering}@{}|@{}}*{1}{m{1.7cm}<{\centering}@{}|@{}}*{1}{m{1.7cm}<{\centering}@{}|@{}}*{1}{m{1.7cm}<{\centering}@{}|@{}}*{1}{m{1.7cm}<{\centering}@{}|@{}}}
				\hline
				 & \multicolumn{3}{c|}{Lesion segmentation} & \multicolumn{2}{c|}{RECIST estimation} \\ \cline{2-6}
                \multirow{-2}{*}{Method} & Precision & Recall & Dice & Long axis & Short axis \\ \hline
				\hline
				Cai \etal \cite{cai2018accurate}  &  0.893$\pm$0.111  &  0.933$\pm$0.095 & 0.906$\pm$0.089  & -  & -  \\ \hline
				Tang \etal \cite{tang2018semi}  &  - & - & - & 1.893$\pm$2.185 &  1.614$\pm$1.874  \\ \hline
				MR-CNN  & 0.850$\pm$0.115  & 0.820$\pm$0.108  & 0.827$\pm$0.092  & 2.361$\pm$2.878  &  1.983$\pm$2.293  \\ \hline
				MR-CNN+PSPNet \cite{zhao2017pyramid}  & 0.893$\pm$0.057 & 0.906$\pm$0.121  & 0.887$\pm$0.078  & 2.057$\pm$2.215  &  1.730$\pm$2.029   \\ \hline
				MR-CNN+DeepLabv3+ \cite{chen2018encoder} & \textbf{0.909$\pm$0.051} & 0.896$\pm$0.127  & 0.891$\pm$0.067  & 2.023$\pm$2.243  &  1.715$\pm$2.007  \\ \hline
				MR-CNN+U-Net \cite{ronneberger2015u}  & 0.877$\pm$0.076  &  0.881$\pm$0.143 &  0.867$\pm$0.084 & 2.153$\pm$2.537  &  1.848$\pm$2.153  \\ \hline
				MR-CNN+ResNet \cite{xiao2018simple} & 0.907$\pm$0.047  & 0.875$\pm$0.132  & 0.883$\pm$0.077  & 2.076$\pm$2.311  & 1.742$\pm$2.041   \\ \hline
				MR-CNN+RU-Net  & 0.891$\pm$0.053  &  0.913$\pm$0.113 & 0.901$\pm$0.063  & 1.913$\pm$2.163  &  1.631$\pm$1.911  \\ \hline
				MR-CNN+ARU-Net$^\star$  & 0.879$\pm$0.061  &  0.935$\pm$0.084 &  0.907$\pm$0.051 &  1.908$\pm$2.089 & 1.608$\pm$1.902   \\ \hline
				MR-CNN+ARU-Net  & 0.883$\pm$0.057  &  \textbf{0.947$\pm$0.074} &  \textbf{0.912$\pm$0.039} &  \textbf{1.747$\pm$1.983} & \textbf{1.555$\pm$1.808}   \\ \hline
			\end{tabular}
		}
	\end{center}
	\vspace*{-0.5\baselineskip}
		\scriptsize
		Note: - denotes the result is not reported. $\star$ denotes ARU-Net uses the LOI directly extracted by MR-CNN as input rather than the adjusted one.
\end{table}

\textbf{Quantitative Lesion Segmentation \& RECIST Estimation Results.}

To investigate the benefits of our designs, we evaluate the following experimental configurations: 1) only using the first component (MR-CNN); 2) using U-Net \cite{ronneberger2015u} as the second component (MR-CNN+U-Net); 3) using a good human pose estimation approach (\ie ResNet-Pose \cite{xiao2018simple}) as the second component (MR-CNN+ResNet); 4) using the designed second component without the ASPP module (MR-CNN+RU-Net); 5) directly using the output of MR-CNN as the input of ARU-Net (MR-CNN+ARU-Net$^\star$); 6) the fully proposed SEENet framework (MR-CNN+ARU-Net). For experimental comparisons, two state-of-the-art image segmentation methods (\ie PSPNet \cite{zhao2017pyramid} and DeepLabv3+ \cite{chen2018encoder}) are used as the second component. Also, the best existing weakly-supervised universal lesion segmentation approach \cite{cai2018accurate} and semi-supervised RECIST estimation approach \cite{tang2018semi} are compared.

Table \ref{tab:recist} lists the quantitative results of lesion segmentation and RECIST estimation produced by different methods. From Table \ref{tab:recist}, we can see that 1) when using the second component for refinement, the performance of both lesion segmentation and RECIST estimation is improved significantly, demonstrating that the importance and effectiveness of the refinement step in the proposed framework. 2) Compared with U-Net and ResNet, RU-Net achieves better results, suggesting that RU-Net combining U-Net and ResNet is able to learn more representative features for performance improvement. 3) ARU-Net outperforms RU-Net, suggesting that the introduced ASPP module can extract multiscale contextual and discriminative features to strengthen the model's capability for lesion segmentation and RECIST estimation. 4) ARU-Net gets better results than ARU-Net$^\star$, especially for RECIST estimation. It demonstrates that the simple adjustment step can reduce the arbitrariness of lesion poses, which makes the model learn more discriminative features for our tasks. 5) Compared with the best approaches \cite{cai2018accurate,tang2018ct,tang2018semi,agarwal2020weakly} customized for each specific task and requiring radiologists to draw bounding boxes to extract lesion of interests (LOIs), the proposed unified framework SEENet (\ie MR-CNN+ARU-Net) still gets better performance on both tasks with less human effort (\ie one click), demonstrating the effectiveness of the proposed method for simultaneous lesion segmentation and RECIST estimation. 6) Compared with the state-of-the-art image segmentation methods (\ie PSPNet \cite{zhao2017pyramid} and DeepLabv3+ \cite{chen2018encoder}), the proposed ARU-Net achieves better results, meaning that the well-designed ARU-Net is more powerful and suitable for our tasks.

\textbf{Quantitative Lesion Detection Results.}
Table \ref{tab:localization} lists the quantitative results of lesion detection using different approaches. As mentioned, when without using the click information, MR-CNN can be used for fully automatic lesion detection, denoted by MR-CNN w/o click. It achieves competitive performance compared with the state-of-the-art lesion detection approaches \cite{yan20183d,li2019mvp,zlocha2019improving,tao2019improving,yan2019mulan}. If the RECIST estimation branch is removed (denoted by MR-CNN w/o click \& RECIST), the detection performance drops about 3\% of sensitivity at 1 false positive per image. It demonstrates the usefulness of the added RECIST estimation branch for lesion detection. When using the click information, MR-CNN achieves a sensitivity of 97.24\% at 0.5 false positives per image, suggesting that the LOIs can be well extracted. We find that some errors happen when there are multiple lesions near the click position.

\begin{table}[t!]
	\begin{center}
		\caption{Given an IoU of 0.5, sensitivity (\%) at several specific average false positives per image on FROC curve. We average the results of MR-CNN with randomly clicking ten times.}
		\label{tab:localization}
		{
			\scriptsize
			
			\begin{tabular}{|@{}*{1}{m{4.5cm}<{\centering}@{}|@{}}*{6}{m{1.25cm}<{\centering}@{}|@{}}}
				\hline
				Method & 0.5 & 1 & 2 & 4 & 8 & 16\\ 
				\hline
				\hline
				3DCE \cite{yan20183d}  &  62.48 & 73.37  &  80.70 & 85.65  &  89.09 & 91.06\\ \hline
				3DCE\_CS\_Att \cite{tao2019improving}  &  71.4 & 78.5  &  84.0 & 87.6  &  90.2 & 91.4 \\ \hline
				Zlocha et al. \cite{zlocha2019improving}  &  72.15 & 80.07  &  86.40 & 90.77  &  94.09 & 96.32 \\ \hline
				MVP-Net \cite{li2019mvp}  &  73.83 & 81.82  &  87.60 & 89.57  &  91.30 & - \\ \hline
				MULAN \cite{yan2019mulan}  &  \textbf{76.12} & 83.69  &  \textbf{88.76} & \textbf{92.30}  &  94.71 & \textbf{95.64} \\ \hline
				MR-CNN w/o click \& RECIST &  72.81 & 80.72  &  86.87 & 91.15  &  94.18 & 95.24\\ \hline
				MR-CNN w/o click  &  75.92 & \textbf{83.74}  &  88.13 & 92.11  &  \textbf{94.82} & 95.63\\ \hline
				\hline
				MR-CNN  &  97.24 &  98.37 & 99.31  & 99.47  & 99.47  & 99.47\\ \hline
			\end{tabular}
		}
	\end{center}
	
\end{table}

\section{Conclusions and Future Work}
We propose SEENet for automatic lesion segmentation and RECIST estimation on a variety of lesion types and require very simple human guide information, \ie just one click. To obtain reliable lesion of interests, an improved Mask R-CNN model is designed by adding an extra two-channel input and a new branch for RECIST estimation. To further boost the performance of lesion segmentation and RECIST estimation, ARU-Net is designed to learn discriminative features with multiscale contextual information. Experimental results demonstrate the proposed method can work well on behalf of ``click in and lesion measurements out'' for clinicians. As such, it shows highly potential clinical values. Future work includes extending SEENet to 3D and making it end-to-end trainable.

\noindent\textbf{Acknowledgments.}
This research was supported by the Intramural Research Program of the National Institutes of Health Clinical Center and by the Ping An Insurance Company through a Cooperative Research and Development Agreement. We thank Nvidia for GPU card donation.

\bibliographystyle{splncs04}
\bibliography{egbib}
\end{document}